\documentclass[preprint,showpacs,preprintnumbers,amsmath,amssymb,pre]{revtex4}

\usepackage{pifont,graphicx}

\usepackage{dcolumn}
\usepackage{bm}

\begin{document}


\title{Dynamical and Stationary Properties of On-line Learning from Finite 
Training Sets}
\date{\today}

\author{Peixun Luo}
 \email{physlpx@ust.hk}
\author{K. Y. Michael Wong}
 \email{phkywong@ust.hk}
\affiliation{
Department of Physics, Hong Kong University of Science and Technology,\\ 
Clear Water Bay, Kowloon, Hong Kong
}


\begin{abstract}
The dynamical and stationary properties of on-line learning 
from finite training sets are analysed using the cavity method. 
For large input dimensions, 
we derive equations for the macroscopic parameters, namely,
the student-teacher correlation, the student-student autocorrelation
and the learning force fluctuation.
This enables us to provide analytical solutions to Adaline learning as a
benchmark. Theoretical predictions of training errors in transient 
and stationary states are obtained by a Monte Carlo sampling procedure.
Generalization and training errors are found to agree with simulations. 
The physical origin of the critical learning rate is presented.
Comparison with batch learning is discussed throughout the paper.
\end{abstract}


\maketitle

\section{\label{intro}Introduction}

In recent years, there have been many attempts to analyse the 
dynamics of learning from examples in classification and regression 
\cite{Mace:Coolen}.
This refers to the dynamical process
of minimizing the risk functions of the classifier or regressor, 
often via gradient descent, until a steady state is reached. 
Despite progress in understanding
the {\it steady-state} behavior of learning processes \cite{Bishop},
the {\it dynamics} of learning was much less understood.
This is probably due to the high complexity in its analysis,
since it typically involves the evolution of many microscopic parameters,
each strongly interacting with others in a convolutional way.
Yet, a number of important issues in improving the learning efficiency
depend on a better understanding of its dynamics,
including the speed of convergence,
the early stopping point for optimal generalization,
the shortening of the plateau regime,
and the avoidance of getting trapped in local minima
\cite{Saad,Wong:Li:Tong,EPL}.
Hence, it would be both useful and challenging
to analyse the dynamics of learning.

{\it On-line learning} is a common mode of implementing learning,
in which an independent example is presented at each learning step.
Significant progress has been made 
in the case of on-line learning of {\it infinite} training sets 
\cite{biehl-schwarze,Saad-Solla,Saad}.
Since statistical correlations among the examples can be ignored,
the dynamics can be described by instantaneous dynamical variables,
leading to great advances in our understanding of on-line learning.
However, in reality, the same restricted set of examples
is recycled during the learning process.
This introduces temporal correlations of the weights in the learning history,
rendering the analysis at best an approximation to the reality.

There were some attempts to understand on-line learning with {\it recycled} 
examples. Early researchers used the approximate Fokker-Planck equation to 
describe the learning process \cite{Radons,Hansen}. The use of perturbative 
expansions of the master equation was shown to be insufficient for a precise 
calculation of global properties of on-line learning \cite{Heskes}.  
The difference between batch learning and on-line learning was investigated 
to the first order of the learning rate \cite{Heskes-kappen}. For general 
learning rates, the exact solution for Hebbian rule was derived in 
Ref.~\cite{Rae:Sollich:Coolen}. Exact solutions were found for linear networks,
and the generalization ability of on-line learning was found to 
outperform batch learning if bias is present in the input 
\cite{Barber:Sollich}. The dynamics of on-line learning in multilayer neural 
networks were analysed using the dynamical replica method and solutions were
found in the limit of large sizes of training sets \cite{Coolen:Saad:Xiong}.

A recent work based on the generating functional approach is a good step 
forward toward a general theory of describing the dynamical and stationary 
properties of on-line learning \cite{Heimel:Coolen}. It illustrates the 
mean-field character of the dynamics in its description in terms of an 
effective single example. For random choices of the sequence of presented 
examples, the dynamics is characterized by the appearance of an example
as a Poisson event in the learning sequence. Steady state properties were 
discussed by neglecting fluctuations in the learning forces 
(referred to as the mean-force approximation hereafter).

In this paper, we propose an analysis of on-line learning with recycled 
examples using the cavity method. The cavity method is a mean-field analysis 
first used in magnetic systems \cite{MPV}.
It enables us to understand the properties of a system
by focusing on the response of the system to a single element added to it.
It was later generalized to study learning in neural networks 
with the advantages of a clear physical picture and 
microscopic insights to both their equilibrium and dynamical properties
\cite{Luo-wong,EPL,Wong:Li:Tong}. The cavity method was subsequently applied 
to analyse the dynamics of {\it batch} learning, in which the entire set of 
examples is provided for each learning step \cite{Wong:Li:Tong}, It provides
dynamical equations and obtains important results on the overtraining, early 
stopping, noise effects and average learning strategy.

To adapt the cavity method from batch learning to on-line learning in this 
paper, there is a need to account for the following subtleties.
(a) Averaging over the choice of sequencing the examples is now necessary.
(b) The measurements of an example observed at an instant
is now correlated with the instants when it was learned.
This is due to the giant boost of that example at a learning step,
which upsets the uniformity of the examples
as in the case of batch learning.

The purposes of this paper are:
(a) to perform an exact analysis of the learning dynamics
as far as the formulation allows,
so that minimal approximations are made,
and deeper physical insights can be extracted;
(b) to illustrate the analytical approach
using the simple example of a linear learning rule,
which can act as a benchmark for verifying the validity of the theory,
and a theoretical framework for more complicated systems,
such as nonlinear learning rules and multilayer networks;
(c) to explore efficient Monte Carlo procedures
implied by the distribution of example activations predicted by the theory, 
which can be applied to the more complicated cases;
(d) to study the difference between on-line learning and batch learning for 
general learning conditions.

The paper is organized as follows. In Sec.~II we describe the dynamics of 
on-line learning. In Sec.~III we introduce the cavity method and derive the
dynamical equations for the macroscopic measurements: (a) $G(t,s)$ and 
$D(t,s)$, the Green's functions of weights and examples in response to stimuli;
(b) $R(t)$, the correlation between the teacher and student weight vectors,
and $C(t,s)$, the autocorrelation between student weight vectors at different
times; (c) the fluctuation of the learning force $\langle F^2(t)\rangle$. 
The Monte Carlo procedure to calculate the training error is also presented. 
In Sec.~IV, we compare theoretical predictions with simulation results. 
The average learning strategy in the long time limit is 
proposed and compared with the performance of batch learning. 
In Sec.~V, we summarize our work and propose some future directions. 
In Appendix, we describe the mathematical details of the procedure of sequence
averaging.

\section{Formulation}
\label{sec:formu}

We consider a training set of $p$ examples
generated by a teacher network with $N$ weights $B_j$, $j=1,\cdots,N$. 
For definiteness, we set $|{\boldsymbol B}|=1$.
Each example $\mu=1,\cdots,p$ consists of an $N$-dimensional input vector
$\boldsymbol\xi^\mu$, and a teacher generated output $\tilde y_\mu$.
It is convenient to introduce the parameter $\alpha\equiv p/N$.
The inputs $\xi^\mu_j$ are Gaussian variables
with zero mean and unit variance. The outputs are
\begin{displaymath}
	\tilde y_\mu=\left\{
	\begin{array}{ll}
	{\rm sgn}(y_\mu+\varepsilon z_\mu)&\textrm{for classification}  \\
	y_\mu+\varepsilon z_\mu 	  &\textrm{for regression,}
	\end{array}\right.
\end{displaymath}
where $y_\mu\equiv\boldsymbol B\cdot\boldsymbol\xi^\mu$
is the teacher activation,
$z_\mu$ is a Gaussian variable with zero mean and unit variance,
and $\varepsilon$ is the noise amplitude.

The examples are learned by a student network
with the same number of inputs and output.
At each learning step, one example in the training set is randomly drawn.
If the example drawn out at time $t$ is $\sigma(t)$,
then the weights are modified according to 
\begin{eqnarray} 					\label{J}
       J_j(t+\frac{1}{N})
	&=&J_j(t)+\frac{v}{N}[F\mathbf{(}x_{\sigma(t)}(t),\tilde y_{\sigma(t)}
	\mathbf{)} \xi^{\sigma(t)}_ j-\lambda J_j(t)]+\frac{1}{N}\eta_j(t),
\end{eqnarray}
where $x_{\sigma(t)}(t)\equiv{\boldsymbol J}(t)\cdot\boldsymbol
\xi^{\sigma(t)}$ is the student activation. $v$ is the learning rate and 
$\lambda$ is the weight decay. The force $F(x,\tilde y)$ describes the 
learning rule,
\begin{equation}
	F(x,\tilde y)=
	\begin{cases}
	\tilde y &\textrm{for Hebbian rule}	\\
	\tilde y-x &\textrm{for Adaline rule}	\\
        \tilde y(\kappa-x\tilde y)\Theta(\kappa-x\tilde y)
        &\textrm{for Adatron rule},
	\end{cases}
\end{equation}
where $\Theta$ is the step function and $\kappa$ is the stability.
The last term in Eq.~(\ref{J}) is the dynamical noise term,
often added to avoid the learning procedure being trapped in local minimum,
with $\langle\eta_j(t)\eta_k(s)\rangle=2T\delta_{ts}/N$
and $T$ is the dynamical noise level.

In the limit of vanishing learning rate $v$, 
the on-line dynamics described by Eq.~(\ref{J}) 
is equivalent to the batch learning formulation in \cite{Wong:Li:Tong} 
when the time scale, weight decay and the dynamical noise in the latter are 
multiplied by factors of $\alpha/v$, $1/\alpha$ and $v^2/\alpha^2$ 
respectively. However, for finite learning rate $v$,
the randomness of the learning sequence adds noise to the dynamics.

\section{The Cavity Method}
\label{sec:cavity}

\subsection{The cavity activation and the Green's functions}

Consider a new example 0 that is not included in the original training set. 
We define its activation at time $t$ in the network trained without that 
example as its {\em cavity activation} $h_0(t)$, i.~e., 
$h_0(t)\equiv{\boldsymbol J}(t)\cdot{\boldsymbol\xi^0}$. It
is a random variable since the network has not learned the information
of this new example. When the size of the network $N$ is very large, 
$h_0(t)$ is a Gaussian variable with mean $R(t)y_0$
and covariance $C(t,s)-R(t)R(s)$,
where $R(t)\equiv\boldsymbol B\cdot\boldsymbol J(t)$
is the student-teacher correlation at time $t$,
and $C(t,s)\equiv\boldsymbol J(t)\cdot\boldsymbol J(s)$
is the student-student autocorrelation at times $t$ and $s$.
Both $R(t)$ and $C(t,s)$ are self-averaging in the limit
of large $N$.

Now we consider the evolution of another network $J_j^0(t)$ in which the 
example 0 is added to its training set. To ensure that the probability of 
occurrence of the new example 0 and the old ones remain identical, the new 
example sequence $\sigma^0(t)$ is obtained from the original example sequence 
$\sigma(t)$ according to 
\begin{equation}				\label{new-sq} 
	\sigma^0(t)=
	\begin{cases}
	\sigma(t) \quad	&\textrm{probability}=1-\frac{1}{1+p}	\\
	0	  \quad &\textrm{probability}=\frac{1}{p+1}. 
	\end{cases}			
\end{equation}
In the new example sequence $\sigma^0(t)$, at each 
learning step, the weight is modified according to
\begin{equation}   				\label{J-0}
       J_j^0(t+\frac{1}{N})=J_j^0(t)
	+\frac{v}{N}[F(x_{\sigma^0(t)}^0(t),\tilde y_{\sigma^0(t)})
	\xi^{\sigma^0(t)}_j-\lambda J_j^0(t)]+\frac{1}{N}\eta_j(t) ,   
\end{equation}
where $x_{\sigma^0(t)}^0(t)\equiv{\boldsymbol J}^0(t)
	\cdot{\boldsymbol \xi}^{\sigma^0(t)}$.  
Compared the networks $J_j^0(t)$ and $J_j(t)$, we obtain from Eqs.~(\ref{J})
and (\ref{J-0}),
\begin{equation}
\begin{split}
	({\hat S}-1+\frac{v\lambda}{N})(J_j^0(t)-J_j(t))	
        =\frac{v}{N}[F(x_{\sigma^0(t)}^0(t),{\tilde y}_{\sigma^0(t)}(t))
	\xi_j^{\sigma^0(t)} 	
	   -F(x_{\sigma(t)}(t),{\tilde y}_{\sigma(t)}(t))\xi_j^{\sigma(t)}],
\end{split}
\end{equation}
where $\hat S$ is the time shift operator.
Let $G^{(0)}(t-t')$ be the bare Green's function
\begin{equation}
        G^{(0)}(t-t')=\Theta(t^+-t'-\frac{1}{N})(1-\frac{v\lambda}{N})
	^{N(t-t'-\frac{1}{N})}.
\end{equation}
It satisfies 
\begin{equation}
	(\hat S-1+\frac{v\lambda}{N})G^{(0)}(t-t')=\delta_{tt'}.
\end{equation}
We assume that the adjustment from $J_j(t)$ to $J^0_j(t)$ is small so that
linear response theory is applicable. Then on separating the contributions 
from the new example and the old ones, we have
\begin{widetext}
\begin{eqnarray}				\label{J0-J}
        J_j^0(t)-J_j(t)&=&\frac{v}{N}\sum_{t'<t}G^{(0)}(t-t')\delta_
        {\sigma^0(t')0}[F_0(t')\xi_j^0-F_{\sigma(t')}(t')\xi_j^
	{\sigma(t')}]				\nonumber \\
	&&+\frac{v}{N}\sum_{k,t'<t}G^{(0)}(t-t')\delta_{\sigma^0(t')\sigma(t')}
        \xi_j^{\sigma(t')}F'_{\sigma(t')}(t')\xi_k^{\sigma(t')}
        [J_k^0(t')-J_k(t')],
\end{eqnarray}
\end{widetext}
where $F_{\sigma(t)}(t)$  is the shorthand notation
of the force acting on example $\sigma(t)$ at time $t$,
and $F'$ represents the derivative of the force
with respect to the activation $x$.
We can now interpret this result
from the viewpoint of the linear response theory.
The first term on the right hand side describes the primary effects of
adding example 0 to the training set
and is the driving term for the difference between the two networks.
This occurs at the discrete instants with $\sigma^0(t')=0$
by adding the force due to example 0 and removing that due to
the original example $\sigma(t')$.
The second term describes the many-body
reactions due to the change of the original examples caused by the added 
example, and is referred to as the Onsager reaction term. Describing the 
response to the driving term by the Green's function,
Eq.~(\ref{J0-J}) reduces to
\begin{equation}				\label{J0-J-a}
	J_j^0(t)-J_j(t)=\frac{v}{N}\sum_{k,t'<t}G_{jk}(t,t')
        \delta_{\sigma^0(t')0}[F_0(t')\xi_k^0-F_{\sigma(t')}(t')
	\xi_k^{\sigma(t')}],
\end{equation}
where $G_{jk}(t,s)$ is the time-dependent Green's function with
iterative expression 
\begin{equation}				\label{Green} 
	G_{jk}(t,s)=\delta_{jk}G^{(0)}(t-s)+\frac{v}{N}\sum_{t'<t}\sum_l
        G^{(0)}(t-t')\delta_{\sigma^0(t')\sigma(t')}
        \xi_j^{\sigma(t')}F'_{\sigma(t')}(t')\xi_l^{\sigma(t')}G_{lk}(t',s).
\end{equation}
The Green's function $G_{jk}(t,s)$
is the response of the weight $J_j$ at time $t$
due to a unit stimulus added at time $s$
to the right hand side of Eq.~(\ref{J}) corresponding to weight $J_k$,
in the limit of vanishing magnitude of the stimulus.

In the limit of large $N$, we can apply a diagrammatic analysis similar to the 
case of batch learning \cite{Wong:Li:Tong}. In contrast with batch learning, we
need to first average Eqs.~(\ref{J0-J-a}) and (\ref{Green}) over the 
distribution of example sequence using Eq.~(\ref{new-sq}). This can then be 
followed by the usual averaging over the distribution of background examples, 
as in the case of batch learning. The result is that we can neglect the effect
of removing the background example represented by the second term in the square
bracket of the right hand side of Eq.~(\ref{J0-J-a}). $G_{jk}(t,s)$ is self-averaging and diagonal in larger $N$ limit, so that 
$G_{jk}(t,s)=G(t,s)\delta_{jk}$, where $G(t,s)$ satisfies the Dyson's equations
\begin{eqnarray}
	G(t,s)&=&G^{(0)}(t-s)+v\int{\rm d}t_1\int{\rm d}t_2 G^{(0)}(t-t_1)
	\langle D_{\sigma(t_2)}(t_1,t_2)F'_{\sigma(t_2)}(t_2)\rangle G(t_2,s)  
					\label{green}\\
	D_{\sigma(s)}(t,s)&=&\delta(t-s)+\frac{v}{\alpha}\int{\rm d}t_1
			G(t,t_1)F'_{\sigma(s)}(t_1)D_{\sigma(s)}(t_1,s),
					\label{D}
\end{eqnarray}
where $G^{(0)}(t-s)\equiv \Theta(t-s)\exp[-v\lambda(t-s)]$
is the bare Green's function.
$D_{\sigma(s)}(t,s)$ is the {\em example Green's Function},
and $\langle\,\rangle$ represents average
over distributions of both example sequences and examples. 

We emphasize that the average 
$\langle D_{\sigma(t_2)}(t_1,t_2)F'_{\sigma(t_2)}(t_2)\rangle $
is different from the average 
$\langle D_\mu(t_1,t_2)F'_\mu(t_2)\rangle $. The former specifies that the 
function $F'(t_2)$ and $D(t_1,t_2)$ are due to the example that was picked from
the example sequence for learning at the particular instant $t_2$. During 
on-line learning, the activation of this example receives a giant boost at the
learning instant, as mentioned later in the text discussion of Fig.~1. This 
makes its distribution different from that of a randomly drawn
example $\mu$, whose previous learning instant remains unspecified. Hence, the
former average will be referred to as an {\em active} average, in contrast
to the latter, which is referred to as a {\em passive} average.

Nevertheless, in the case of linear rules used for illustration later in this
paper, $F'(t)$ is a constant independent of $t$. Hence, the active average in 
Eq.~(\ref{green}) becomes identical to the passive average. 
Thus, the Dyson's equations
(\ref{green}) and (\ref{D}) becomes identical to those of batch learning 
\cite{Wong:Li:Tong}, after rescaling the time and the weight decay. 

In the case of Hebbian rule, $F'(x)=0$ and $D_\mu(t,s)=\delta(t-s)$. The 
Green's function becomes identical to the bare Green's function.

In the case
of the Adaline rule, $F'(x)=-1$ and $D_\mu(t,s)=D(t,s)$ independent of example
$\mu$. The weight Green's function becomes invariant under translation of 
time, and can be written as   
\begin{equation}
	G(t,s)=G(t-s,0)=\int\rho(x)e^{-x(t-s)}{\rm d}x,
\end{equation}
where $\rho(x)$ is the density of state
\begin{equation}
	\rho(x)=(1-\alpha)\Theta(1-\alpha)\delta(x-v\lambda)
	+\frac{\sqrt{(x_{\max}-x)(x_{\min}-x)}}{2\pi\frac{v}{\alpha}
	(x-v\lambda)},
\end{equation}
with $x_{\max}$ and $x_{\min}$ are the edges of the spectrum given by 
$x_{\max},x_{\min}=v\lambda+v(1\pm 1/\sqrt{\alpha})^2$, respectively.

The number of times $m$ that the new example 0 appears in time $t$ follows a 
Poisson distribution with mean $t/\alpha$. If these appearances occur
at times $t_1,\cdots, t_m\ (t_m<\cdots<t_1<t)$, Eq.~(\ref{J0-J-a}) reduces to 
\begin{equation}
        J_j^0(t)=J_j(t)
        +\frac{v}{N}\sum_{r=1}^m G(t,t_r)F_0(t_r)\xi^0_j.
\end{equation}
Multiplying both sides by $\xi^0_j$ and summing over $j$, one derives the
relationship between the cavity activation and the generic activation of 
example 0,
\begin{equation}
        x_0(t)=h_0(t)+v\sum_{r=1}^m G(t,t_r)F_0(t_r).
\label{evolve}
\end{equation}
This relation enables us to express the cavity activation $h(t)$ of any 
example as a function of its generic activation 
$x(t_1),\cdots,x(t_m),x(t)$ at the previous and current learning instants,
and attributes physical meaning to the single effective example
in Ref.~\cite{Heimel:Coolen}.
Hereafter, we omit the subscript if no confusion occurs.

\begin{figure}[htb] 
\begin{center}
\includegraphics[width=0.45\textwidth]{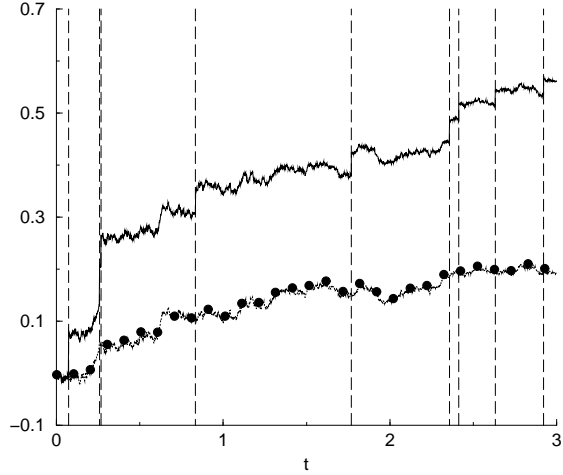}
\caption{The evolution of the activations of a randomly selected example
in a network with $N=1000, \alpha=0.3, \kappa=0.8, v=0.1$ and $\lambda=0.1$
using the Adatron rule.
See text discussions for the explanations of the lines and symbols.}
\label{fig:cav}
\end{center}
\end{figure}

The simulation results in Fig. \ref{fig:cav} verify the relationship between 
the cavity activation and the generic activation
for a randomly selected example.
Up to $t=3$, the example is drawn from the learning sequence $\sigma(t)$ 
9 times, close to the Poisson average of $t/\alpha=10$.
The solid line describes the evolution of $x(t)$,
which exhibits giant boosts at the 9 learning instants
indicated by the vertical dashed lines.
The dotted line describes the evolution of the cavity activation $h(t)$,
which is obtained in a second network
which uses the same learning sequence $\sigma(t)$,
except that learning is paused when the example is drawn.
Since the example and this network are uncorrelated,
$h(t)$ evolves as a random walker with appropriate means and covariances.
The filled circles indicate the values of the cavity activations
predicted by Eq. (\ref{evolve}),
using the Green's functions measured
by comparing learning with and without stimuli \cite{Wong:Li}.
They show remarkable agreement with the simulated $h(t)$.

To derive the distribution of generic activations, we first consider the 
distribution of cavity activations, which is given in the Gaussian form at
$m$ learning steps and time $t_0(=t)$ by 
\begin{equation}						\label{P-h}
	P(h(t_0),\cdots,h(t_m)|y)=  	
	\frac{\exp\{-\frac{1}{2}\sum_{i,j=0}^m
	[h(t_i)-R(t_i)y]
        (C-RR^T)^{-1}_{ij}[h(t_j)-R(t_j)y]\}}
        {\sqrt{(2\pi)^{m+1}\det(C-RR^T)}},
\end{equation}
where $C-RR^T$ is a square matrix with size $m+1$ and 
$(C-RR^T)_{ij}=C(t_i,t_j)-R(t_i)R(t_j)$.
The corresponding distribution of generic activations can be written as
\begin{equation}
	P(x(t_0),\cdots,x(t_m)|y,\tilde y)
	=P(h(t_0),\cdots,h(t_m)|y)
	\left|\frac{\partial(h(t_0),\cdots,h(t_m))}
	{\partial(x(t_0),\cdots,x(t_m))}\right|,
\end{equation}
where $h(t_i)$ is a function of $x(t_{i}),\cdots,x(t_{m})$ defined by
Eq.~(\ref{evolve}), and the dependence on $\tilde y$ may arise from the 
learning forces. Since $\partial h(t_i)/\partial x(t_j)=0$ for 
$t_j>t_i$, and $\partial h(t_i)/\partial x(t_i)=1$, the Jacobian reduces 
to 1. Therefore, the distribution of generic activations can be 
expressed as
\begin{equation}					\label{P-x}
	P(x(t_0),\cdots,x(t_m)|y,\tilde y)=  	
	\frac{\exp\{-\frac{1}{2}\sum_{i,j=0}^m [h(t_i)-R(t_i)y]
	(C-RR^T)^{-1}_{ij}[h(t_j)-R(t_j)y]\}}
        {\sqrt{(2\pi)^{m+1}\det(C-RR^T)}}.
\end{equation}
In general, $h(t_i)$ can be a nonlinear function of 
$x(t_i),\cdots,x(t_m)$. Hence, the generic activation distribution in 
Eq.~(\ref{P-x}) is no longer Gaussian, although the cavity activation 
distribution in Eq.~(\ref{P-h}) is. This characteristic of on-line learning is 
demonstrated in numerical simulation for Adatron rule in 
Ref.~\cite{Coolen:Saad}. 

We now illustrate how the above result can be applied to specific cases. For
Hebbian rule, Eq.~(\ref{evolve}) implies that $h(t)$ is not an explicit 
function of $x(t_r)$ at the previous learning instants $t_r$,
\begin{equation}
        h(t)=x(t)-\tilde y\sum_{r=1}^m\exp[-v\lambda(t-t_r)].
\end{equation}
This enables us to write down the instantaneous activation distribution, given
the learning instants $t_1,\cdots,t_m$ of the example,
\begin{equation}
	P(x,t_0|y,\tilde y;t_1,\cdots,t_m)
	=\frac{\exp\{-\frac{1}{2}(C(t,t)-R^2(t))^{-1}
	[x(t_j)-\tilde y\sum_{r=1}^m e^{-v\lambda(t-t_r)}-R(t)y]^2\}}
	{\sqrt{2\pi[C(t,t)-R^2(t)]}}.
\end{equation}
The distribution is then averaged over the time distribution and the Poisson
distribution of learning instants
\begin{equation}
	P(x,t|y,\tilde y)=
        \langle P(x,t|y,\tilde y;t_1,\cdots,t_m)\rangle_\sigma.
\end{equation}
The sequence average of an instantaneous quantity $\psi$ at time $t$
depending on the previous learning instants $t_1,\cdots,t_m$ is
\begin{equation}		
\label{seq-avg}
       \langle\psi(t|t_1,\cdots,t_m)\rangle_\sigma=\sum_{m=0}^\infty
        \frac{e^{-\frac{t}{\alpha}}}{\alpha^m}\int_0^t\!{\rm d}t_1\cdots
        \int_0^{t_{m-1}}\!{\rm d}t_m\psi(t|t_1,\cdots,t_m),
\end{equation}  
where the factor of $m!$ in the Poisson distribution
is cancelled by the number of permutations in ordering $t_1,\cdots,t_m$.
Using the Hubbard-Stratonovich identity, we can factorize the integrals over
$t_r (1\leq r\leq m)$. We arrive at the result 
\begin{eqnarray}
	P(x,t|y,\tilde y)&=&\int\frac{{\rm d}\hat x}{2\pi}
        \exp\Big\{i\hat x[x-R(t)y]
	-\frac{1}{2}[C(t,t)-R^2(t)]\hat x^2	\nonumber\\
	&&\phantom{\int\frac{{\rm d}\hat x}{2\pi}\exp\Big\{}
	+\frac{1}{\alpha}\int_0^t{\rm d}s
	[\exp(-i\hat x\tilde ye^{-v\lambda(t-s)})-1]\Big\}, 
\end{eqnarray} 
which agrees with the rule-specific derivation in 
Ref.~\cite{Rae:Sollich:Coolen}.

For Adaline rule, substituting $F(x,\tilde y)=\tilde y-x$ into 
Eq.~(\ref{evolve}) yields a linear relation between the student activation and 
the cavity ones, 
\begin{equation}
        x(t_0)=\sum_{r=0}^m(1+vG)^{-1}_{0r}
	\left[h(t_r)+v\sum_{s=r+1}^mG_{rs}\tilde y\right],
\end{equation}
where $G$ is a square matrix with size $m+1$ and $G_{rs}=G(t_r-t_s,0)$ for 
$t_r>t_s$ and $G_{rs}=0$ for $t_r\leq t_s$. Inserting the mean and variance 
of the cavity activation, we see that $x(t_0)$ is a Gaussian variable with
mean and variance   
\begin{eqnarray}
\label{x-m-v}
	\langle x(t_0)\rangle&=&\sum_{r=0}^m(1+vG)^{-1}_{0r}
	\left[R(t_r)y+v\sum_{s=r+1}^mG_{rs}\tilde y\right] \nonumber\\
	\Delta^2(t_0)&=&\sum_{r,s=0}^m(1+vG)_{0r}^{-1}(1+vG)_{0s}^{-1}
	[C(t_r,t_s)-R(t_r)R(t_s)].
\end{eqnarray}
To obtain the activation distribution in such an application as the average 
training error, we need to further average the Gaussian distribution in 
Eq.~(\ref{x-m-v}) over the learning sequences.

In general, for nonlinear learning rules, the linear inversion of 
Eq.~(\ref{evolve}) to obtain the student activation is not possible, and the 
activation distribution becomes non-Gaussian, even for a given sequence of 
learning instants. Nevertheless, a useful identity exists for the sequence 
average pertaining to an example, as derived in Appendix A,
\begin{equation}                                     \label{x-h}
   \langle x(t)\rangle_\sigma=h(t)+\frac{v}{\alpha}\int_0^t dt'G(t,t')
   \langle F(t')\rangle_\sigma.
\end{equation}
This equation of the sequence-averaged activation 
is the same as that of the self-consistent equation of the activation in batch 
learning, after rescaling the time and weight decay \cite{Wong:Li:Tong}.

\subsection{The student-teacher correlation}

To analyse the student-teacher correlation,
we multiply both sides of Eq. (\ref{J}) by $B_j$ and sum over $j$, 
yielding in the limit of large $N$,
\begin{equation}
    \left(\frac{d}{dt}+v\lambda\right)R(t)=v\langle\langle y_\mu F_\mu(t)
\rangle_\sigma\rangle_\mu,
\label{R}
\end{equation}
where $\langle\cdot\rangle_\mu$ 
represents averaging over the distribution of examples.

For Adaline rule, the solution of $R(t)$ in Eq.~(\ref{R}) involves 
$\langle x_\mu(t)\rangle_\sigma$. By virtue of Eq.~(\ref{x-h}),
and exploiting the example Green's function in Eq.~(\ref{D}), we obtain
\begin{equation}
\label{x-d}
	\langle x_\mu(t)\rangle_\sigma=\int_0^t{\rm d}t_1D(t-t_1)
	\left[h_\mu(t_1)+\frac{v}{\alpha}\int_0^{t_1}{\rm d}t_2G(t_1-t_2)
	\tilde y_\mu\right].
\end{equation}
Applying Laplace transform to Eq.~(\ref{x-d}) and then Eq.~(\ref{R}),
\begin{eqnarray}
	\langle\tilde x_\mu(z)\rangle_\sigma&=&\tilde D(z)	
        \left[\tilde h_\mu(z)+\frac{v\tilde y_\mu}{\alpha z}
		\tilde G(z)\right],
							\nonumber\\
	(z+v\lambda)\tilde R(z)&=&v\left\{\frac{a}{z}-\tilde D(z)
	\left[\tilde R(z)+
		\frac{va}{\alpha z}\tilde G(z)\right]\right\},
\end{eqnarray}
where 
\begin{displaymath}
	a\equiv\langle y_\mu\tilde y_\mu\rangle_\mu=\left\{
	\begin{array}{ll}
        \sqrt{\frac{2}{\pi(1+\varepsilon^2)}}
        & \quad\textrm{for classification},
				  \\
	\sqrt{1+\varepsilon^2}& \quad\textrm{for regression}.
	\end{array}\right.
\end{displaymath}
Here $\tilde G(z)\equiv\int_0^\infty{\rm d}te^{-zt}G(t,0)$ and 
$\tilde D(z)\equiv\int_0^\infty{\rm d}t e^{-zt}D(t,0)$.
Inverse Laplace transform yields
\begin{equation}
     R(t)=a\int{\rm d}x\rho(x)(x-v\lambda)
	\frac{1-e^{-xt}}{x},
\end{equation}
which is the same as that in batch learning after rescaling.

\subsection{The student-student autocorrelation and the force fluctuation}

To analyse the student-student autocorrelation, 
we multiply both sides of Eq. (\ref{J}) by $J_j(s)$ and sum over $j$, 
thus obtaining in the limit of large $N$,
\begin{equation}
      \left(\frac{d}{dt}+v\lambda\right)C(t,s)=v\big\langle\langle x_\mu(s) 
       F_\mu(t)\rangle_\sigma\bigr|_{\sigma(t)=\mu}\big\rangle_\mu.
\label{Cts}
\end{equation}
We remark that $\langle x_\mu(s)F_\mu(t)\rangle_\sigma|_{\sigma(t)=\mu}$,
is an active average, which is distinct from the {\it passive average} 
$\langle x_\mu(s)F_\mu(t)\rangle_\sigma$,
where the example learned at time $t$ is not necessarily $\mu$.

For Adaline learning rule, the average difference  
	$W_\mu(s,t)\equiv\langle x_\mu(s)F_\mu(t)\rangle_\sigma
	|_{\sigma(t)=\mu}-\langle x_\mu(s)F_\mu(t)\rangle_\sigma$ 
can be expressed in terms of Green's functions and learning force, 
\begin{equation} \label{act-pass}
        W_\mu(s,t)=v\int_t^s{\rm d}t'D(s,t')G(t',t)
	\langle F_\mu^2(t)\rangle_\sigma,
\end{equation}
as shown in Appendix \ref{sec:w-c}.
A similar equation for the passive average is also derived therein,
\begin{eqnarray}			\label{passive} 
       \langle x_\mu(s)F_\mu(t)\rangle_\sigma
        &=&\int_0^s{\rm d}t_1 D(s,t_1)
        \left[h_\mu(t_1)\langle F_\mu(t)\rangle_\sigma
        +\frac{v}{\alpha}\int_0^{t_1}{\rm d}t_2 G(t_1,t_2)\tilde y_\mu
        \langle F_\mu(t)\rangle_\sigma\right] \nonumber \\
       && -\frac{v^2}{\alpha}\int_0^{\min(t,s)}dt_1
        \int_{t_1}^s{\rm d}t_2
        \int_{t_1}^t{\rm d}t_3D(s,t_2)G(t_2,t_1)
        D(t,t_3)G(t_3,t_1)\langle F^2_\mu(t_1)\rangle_\sigma. \nonumber\\
\end{eqnarray}
Therefore, one can perform the Laplace transforms 
\begin{eqnarray*}
       \tilde C(w,z)&\equiv&\int_0^\infty{\rm d}s \int_0^\infty{\rm d}t 
	e^{-ws-zt}C(s,t).		\\
        \big\langle\langle\tilde F_\mu(w)\tilde F_\mu(z)
        \rangle_\sigma\big\rangle_\mu
	&\equiv&\int_0^\infty{\rm d}s\int_0^\infty{\rm d}t e^{-ws-zt}
        \big\langle \langle F_\mu(s)F_\mu(t)\rangle_\sigma\big\rangle_\mu. 
\end{eqnarray*}
After substituting Eqs.~(\ref{act-pass}) and (\ref{passive})
into Eq.~(\ref{Cts}), and performing elaborate algebra,
one obtains an equation of the weight correlation
\begin{eqnarray}				\label{C-F}
	\tilde C(w,z)&=&\left[\frac{v^2a^2[\tilde D(w)+\tilde D(z)-1]}{wz}
        +\frac{v^2b}{\alpha zw}
        +v^2\big\langle\langle\tilde F_\mu^2(w+z)\rangle_\sigma\big\rangle_\mu
        +\frac{2T}{(w+z)\tilde D(w)\tilde D(z)}\right]
				 \nonumber \\
	&&\times\frac{\tilde G(z)\tilde D(z)-\tilde G(w)\tilde D(w)}{w-z},
\end{eqnarray}
and an equation of the force autocorrelation
\begin{eqnarray}						\label{F-C}
        \big\langle\langle\tilde F_\mu(w)\tilde F_\mu(z)
        \rangle_\sigma\big\rangle_\mu
        &=&\frac{\tilde D(w)\tilde D(z)}{wz}
        -\frac{va^2}{wz}\tilde G(w)\tilde D^2(w)\tilde D(z)
        -\frac{va^2}{wz}\tilde G(z)\tilde D(w)\tilde D^2(z) \nonumber\\
        &&+\tilde D(z)\tilde D(w)\tilde C(w,z)  
        +\frac{v^2}{\alpha}\tilde G(w)\tilde G(z)\tilde D(w)\tilde D(z)
        \big\langle\langle\tilde F_\mu^2(w+z)\rangle_\sigma\big\rangle_\mu.
					\nonumber\\
\end{eqnarray}
Here 
\begin{displaymath}
	b\equiv\langle\tilde y^2_\mu\rangle_\mu=\left\{
	\begin{array}{ll}
	1		&	\textrm{for classification}  \\
	1+\varepsilon^2	& 	\textrm{for regression.}
	\end{array}\right.
\end{displaymath}

We note the presence of the force fluctuation term  
$\big\langle\langle \tilde F^2(w+z)\rangle_\sigma\big\rangle_\mu$ 
in Eqs.~(\ref{C-F}) and (\ref{F-C}). 
This term is absent in the corresponding equations 
in the case of batch learning. 
This can be seen by observing the scalings $z^{-1} \sim \tilde G(z) \sim
\big\langle\langle\tilde F_\mu^2(w+z)\rangle_\sigma\big\rangle_\mu \sim v^{-1}$
in Eqs.~(\ref{C-F}) and (\ref{F-C}), so that the coupling
of weight correlation $\tilde C(w,z)$ and force autocorrelation 
$\big\langle\langle\tilde F_\mu(w)\tilde F_\mu(z)
\rangle_\sigma\big\rangle_\mu$
via the force fluctuation 
$\big\langle\langle\tilde F_\mu^2(w+z)\rangle_\sigma\big\rangle_\mu$
will approach zero when $v$ is vanishingly small,
which indicates that this temporal correlation is unique to on-line learning.
In contrast to batch learning, the 
presence of the force fluctuation term increases the weight correlation via
Eq.~(\ref{C-F}), which in term increases the force fluctuation itself
via Eq.~(\ref{F-C}). As we shall see, this coupling leads to a collective 
relaxation mode.

Performing the inverse Laplace transform, one can obtain the autocorrelation
\begin{eqnarray}
\label{C}
       C(s,t)&=&\int{\rm d}x \rho(x)(x-v\lambda)\left[\frac{bv}{\alpha}+
        a^2\left(x-v\lambda-\frac{v}{\alpha}\right)\right]
	\left(\frac{1-e^{-xs}}{x}\right)\left(\frac{1-e^{-xt}}{x}\right)
						\nonumber\\
	&&+v\int{\rm d}x\rho(x)(x-v\lambda)
       \int_0^{\min(t,s)}{\rm d}t'e^{-x(t+s-2t')}\langle F^2(t')\rangle 
				\nonumber\\
	&&+T\int{\rm d}x\rho(x)\frac{e^{-x|t-s|}-e^{-x(t+s)}}{x}, 
\end{eqnarray}
where $\langle F^2(t)\rangle\equiv\langle\langle F^2_\mu(t)
\rangle_\sigma\rangle_\mu$. In Eq.~(\ref{C}), the first and third terms are 
similar to those in the autocorrelation function in batch learning. The second 
term represents the contribution induced by the force fluctuations
arising from on-line learning.
It vanishes when $v\to 0$. However, for finite learning rate
$v$, one can see that this term plays a similar role as the dynamic noise.
This can be seen by considering the asymptotic limit of the second term, where 
$\langle F^2(t')\rangle$ approaches the steady-state value of 
$\langle F^2\rangle$, yielding 
\begin{equation}
	\frac{v}{2}\langle F^2\rangle\int{\rm d}x\rho(x)(x-v\lambda)
        \frac{e^{-x|t-s|}-e^{-x(t+s)}}{x}.
\end{equation}
Comparing with the dynamical noise in the third term, we see that 
$\langle F^2\rangle$ is a measure of an effective temperature, and the two 
noise contributions different slightly in their spectrum of relaxation rates.
Therefore, in practice, force fluctuations
can also assist the learning dynamics
in avoiding being trapped in metastable states, playing the same role as 
dynamical noises to batch learning. Hereafter, we let $T=0$ in our final 
results.

To obtain the force autocorrelation 
$\big\langle\langle F_\mu(s)F_\mu(t)\rangle_\sigma\big\rangle_\mu$,
we substitute Eq.~(\ref{C-F}) into Eq.~(\ref{F-C})
and perform the inverse Laplace transform, which yields,
\begin{eqnarray}
        \big\langle\langle F_\mu(t)F_\mu(s)\rangle_\sigma\big\rangle_\mu
        &=&b+\frac{1}{v}\int{\rm d}x\rho(x)\left[\frac{bv}{\alpha}
	+a^2\left(x-v\lambda-\frac{v}{\alpha}\right)\right]
        \Bigg[(x-v\lambda)^2
        \left(\frac{1-e^{-xt}}{x}\right)
	\nonumber\\
	&&\quad\quad\left(\frac{1-e^{-xs}}{x}\right)
	-(x-v\lambda)\left(\frac{1-e^{-xt}}{x}\right)
        -(x-v\lambda)\left(\frac{1-e^{-xs}}{x}\right)\Bigg]    \nonumber \\
        &&+\int_0^{\min(t,s)}{\rm d}t'\langle F^2(t')\rangle
	\int{\rm d}x\rho(x)(x-v\lambda)^2e^{-x(t+s-2t')}.
\end{eqnarray}
Equating $t$ and $s$, the force fluctuation is given by the inverse Laplace 
transform of 
\begin{equation} 					\label{F2}
 	\big\langle\langle\tilde F^2_\mu(z)\rangle_\sigma\big\rangle_\mu
	=\frac{\frac{b}{z}+\frac{1}{v}\int{\rm d}x\rho(x)\frac{x-v\lambda}
	{x^2}\left[\frac{bv}{\alpha}+a^2\left(x-v\lambda-\frac{v}{\alpha}
	\right)\right]\left[\frac{x-v\lambda}{z+2x}+\frac{2v\lambda}{z+x}-
	\frac{x+v\lambda}{z}\right]}{1-\int{\rm d}x\rho(x)
	\frac{(x-v\lambda)^2}{z+2x}}.
\end{equation}
The final expression consists of four contributions. First, the pole at $z=0$ 
gives the steady state value of 
\begin{equation}					\label{F-2}
	\langle F^2\rangle=\frac{b-\frac{1}{v}\int{\rm d}x\rho(x)\left(1-
	\frac{v^2\lambda^2}{x^2}\right)\left[\frac{bv}{\alpha}
        +a^2\left(x-v\lambda-\frac{v}{\alpha}\right)\right]}{1-\int{\rm d}x
	\rho(x)\frac{(x-v\lambda)^2}{2x}}.
\end{equation}
The second and third contributions come form the relaxation spectrum of force 
fluctuations ranging through 
$(x_{\min},x_{\max})$ and $(2x_{\min},2x_{\max})$, 
respectively. The fourth contribution is described by the existence of a 
collective relaxation mode arising from the force-weight coupling, which is
a novel phenomenon of on-line learning. Its relaxation rate $\lambda_\sigma$
is given by the pole 
\begin{equation}
	1-\int{\rm d}x\rho(x)
	\frac{(x-v\lambda)^2}{2x-\lambda_\sigma}=0.
\end{equation} 
This is called slow mode in Ref.~\cite{Barber:Sollich}. When $\lambda_\sigma$ 
approaches zero, the steady-state force fluctuation and student weight will 
diverge. The critical learning rate at which the weight diverges in given
by
\begin{equation}
	v_c=\frac{4}{2-\lambda[1+\alpha+\alpha\lambda 
	-\sqrt{(1+\alpha+\alpha\lambda)^2-4\alpha}]},
\end{equation} 
which is also derived through the spectral analysis \cite{Barber:Sollich}.

\subsection{The training and generalization errors}

The performance of learning is measured by the training and generalization 
errors. Here we provide their expressions of noiseless example. Expression for
other cases can be derived similarly.
In classification, the generalization error is defined as the 
probability that a {\it new} example presented to the network
is misclassified,
$E_g(t)=\langle\Theta[-({\boldsymbol B}\cdot\boldsymbol\xi)
(\boldsymbol J(t)\cdot\boldsymbol\xi)]\rangle_{\boldsymbol\xi}$.
It is determined by the magnitude of the student vector $C(t,t)$ and 
its correlation with the teacher vector $R(t)$ \cite{Wong:Li:Tong}, 
that is, 
\begin{equation}
E_g(t)=\frac{1}{\pi}\cos^{-1}\frac{R(t)}{{\sqrt{C(t,t)}}}.
\end{equation} 
Analytical expressions of $R(t)$ and $C(t,t)$ are derived in previous 
subsections for Adaline rule.
 
The training error is defined as
the fraction of examples in the training set  
that are classified wrongly, i.e.
\begin{eqnarray}
        E_t(t)=&&\int{\rm d}y\int{\rm d}\tilde y P(y,\tilde y)
        \sum_{m=0}^\infty\frac{e^{-\frac{t}{\alpha}}}{\alpha^m}
        \int_0^t{\rm d}t_1\cdots\int_0^{t_m-1}{\rm d}t_m\int{\rm d}x(t)
	\nonumber\\
	&&P(x(t)|y,\tilde y;t_1,\cdots,t_m)\Theta[-\tilde yx(t)].
\end{eqnarray}
This can be computed by a Monte Carlo sampling procedure,
which has been shown to be free from finite size effects \cite{Eiss-opper}.
For general learning rules, we adopt the procedure with the following steps:

1) For a given training example, generate the teacher activation $y$ and 
$\tilde y$ according to $P(y,\tilde y)$.

2) For time $t_0(=t)$, generate the number of times $m$ the example appears 
in a training sequence from time 0 to $t_0$
according to a Poisson distribution with mean $t_0/\alpha$. 

3) Generate the instants $t_1,\cdots,t_m$ that the example appears in the 
training sequence according to a uniform distribution between 0 and $t_0$, 
with $0<t_m<\cdots<t_1<t_0$.

4) Generate the cavity activations $h(t_r)$, $r=0,\cdots,m$ according to the 
Gaussian distribution with mean $R(t_r)y$ and covariance 
$C(t_r,t_s)-R(t_r)R(t_s)$. This can be carried out by generating the 
independent Gaussian variables $z_{ik}$ $(k=0,\cdots,m)$ with mean 0 and 
variance 1, and transforming them to $h(t_i)$ via
\begin{equation}
	h(t_i)=R(t)y+\sum_{k=i}^mA_{ik}z_k,\quad 0\leq i\leq m
\end{equation} 
and the matrix elements $A_{ik}$ are obtained from the recursion relations
\begin{equation}
	A_{mm}=\sqrt{C(t_m,t_m)-R^2(t_m)}
\end{equation}
and for $1\leq j<i\leq m$, 
\begin{eqnarray}
	A_{m-i,m-j}&=&\frac{C(t_{m-i},t_{m-j})-R(t_{m-i})R(t_{m-j})
	-\sum_{k=1}^{j-1}A_{m-i,m-k}A_{m-j,m-k}}{A_{m-j,m-j}}	\nonumber \\
	A_{m-i,m-i}&=&\left[C(t_{m-i,m-i})-R^2(t_{m-i})
	-\sum_{k=1}^{i-1}A^2_{m-i,m-k}\right]^{\frac{1}{2}}. 
\end{eqnarray}

5) Compute $x(t_i)$ according to Eq.~(\ref{evolve}). This enables us to 
collect samples for the distribution $P(x(t_0)|y,\tilde y)$, and hence estimate
the training error.

6) Steps 1) to 5) are repeated to yield sufficient amount of statistics.

For Adaline learning, step 5) of the Monte Carlo sampling procedure can be 
further simplified by exploiting the Gaussian nature of the generic
activation distribution. Using Eq.~(\ref{x-m-v}) to find the mean 
$\langle x(t_0)\rangle$ and variance $\sigma^2(t_0)$, the contribution of a 
example to the training error is given by 
${\rm erfc}(\langle x(t_0)\rangle\tilde y/\sqrt 2\sigma(t_0))$ .

The above procedure assumes that the Green's function $G(t,s)$ and the 
correlations $R(t)$ and $C(t,s)$ are known {\it a priori}. For general learning
rules, $G(t,s)$ can be obtained by solving the Dyson's equations 
Eqs.~(\ref{green}) and (\ref{D}). Since the equation involves an average over 
the distribution of learning sequence and examples, it can again be obtained 
by a Monte Carlo sampling procedure. Similarly, $R(t)$ and $C(t,s)$ can be 
obtained by solving Eqs.~(\ref{R}) and (\ref{Cts}) by Monte Carlo sampling. 
These will be left for further studies. Here, for the exposition 
of the cavity method, we focus on Adaline learning, where these functions can 
be obtained analytically as described in the previous subsections.

For the purpose of consistency check, one can also obtain these functions
directly from simulations and plug into the Monte Carlo sampling procedure to
check whether the generated activation distribution agrees with simulations.

The proposed Monte Carlo procedure is similar to the effective single pattern
process in Ref.~\cite{Heimel:Coolen}.
The difference lies in the generation of the 
learning instants of an example according to the Poisson distribution. In Ref.~
\cite{Heimel:Coolen}, an individual Poisson number with mean $\Delta/\alpha$ 
is generated for every time increment $\Delta$ in the learning history. Here,
the sampling efficiency is improved by a single Poisson number $m$ with mean
$t_0/\alpha$ and $m$ learning instants with uniform distributions.
For general learning rules,
even if the Green's functions and correlation functions have to
be generated from the Monte Carlo sampling procedure, it is possible to use 
similar efficient samplings. This will be left for further studies. 

While the 
Monte Carlo sampling procedure is useful in studying the transient behavior
of learning, it can also be used to extract the stationary properties. At
a very large observation time $t$, we look back at the learning history of 
an example by the network. Reasonably, only those learning events 
occur recently have detectable contributions
to the distribution of the student activation $x(t)$ in Eq.~(\ref{x-m-v}).
Therefore, we can calculate this
distribution to any desired precision by adding the earlier learning events 
one by one, until certain stopping criteria are satisfied. Since the time 
intervals between successive learning events obey an exponential distribution,
we have
\begin{equation}
        P(x|y,\tilde y)=\lim_{m\to\infty}\lim_{t\to\infty}
        \prod_{r=1}^m\left[\int_0^\infty\frac{{\rm d}s_r}{\alpha}
        e^{-\frac{s_r}{\alpha}}\right]
	P(x,t|y,\tilde y;t-s_1,\cdots,t-\sum_{r=1}^ms_r).
\end{equation} 
For Adaline learning, the distribution
$P(x,t|y,\tilde y;t-s_1,\cdots,t-\sum_{r=1}^m s_r)$ is replaced by
a Gaussian distribution with mean and variance given in Eqs.~(\ref{x-m-v}).
We find that the contribution of earlier events approaches zero very quickly 
as $m$ increases. Thus, we only need to invert small matrices in evaluating 
the training error at steady state.
         
\section{Results and discussions}
\label{sec:resul}

Figure 2(a) shows the transient behavior of the three macroscopic parameters, 
student-teacher correlation, student autocorrelation and force fluctuation 
for typical learning parameters. 
The training error and generalization error are shown in Fig.~2(b). 
The theoretical predictions have an excellent agreement with simulations.

\begin{figure}
\begin{center}
\includegraphics[width=0.45\textwidth]{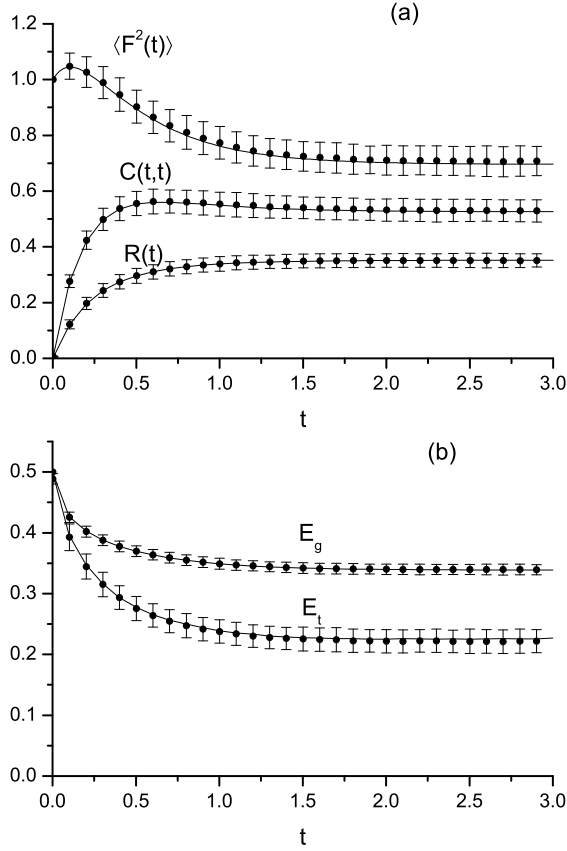}
\caption{The evolution of (a) teacher-student correlation $R(t)$, student 
autocorrelation $C(t,t)$ and force fluctuation $\langle F^2(t)\rangle$, (b) 
training error $E_t$ and generalization error $E_g$ 
for Adaline learning at $\alpha=1.2, v=1.9$, and $\lambda=0.8$.
Solid lines: the cavity method, with analytical results
for $R(t), C(t,t), \langle F^2(t)\rangle, E_g$ and Monte Carlo results for 
$E_t$ averaged over 500,000 samples. 
Symbols: simulations averaged over 100 samples with $N=500$.}
\label{R-C-F-E}
\end{center}
\end{figure}

Figure 3 shows the generalization and training errors at the steady state.
The theoretical predictions agree well with the simulation results. The 
learning dynamics diverges at the critical learning rate $v_c$.
It is also observed that strong weight decays tend to
restrain this divergence at large learning rate,
pushing $v_c$ to higher values.
On the other hand, strong weight decays increase the generalization error
when $v$ is small.

\begin{figure} 
\begin{center}
\includegraphics[width=0.45\textwidth]{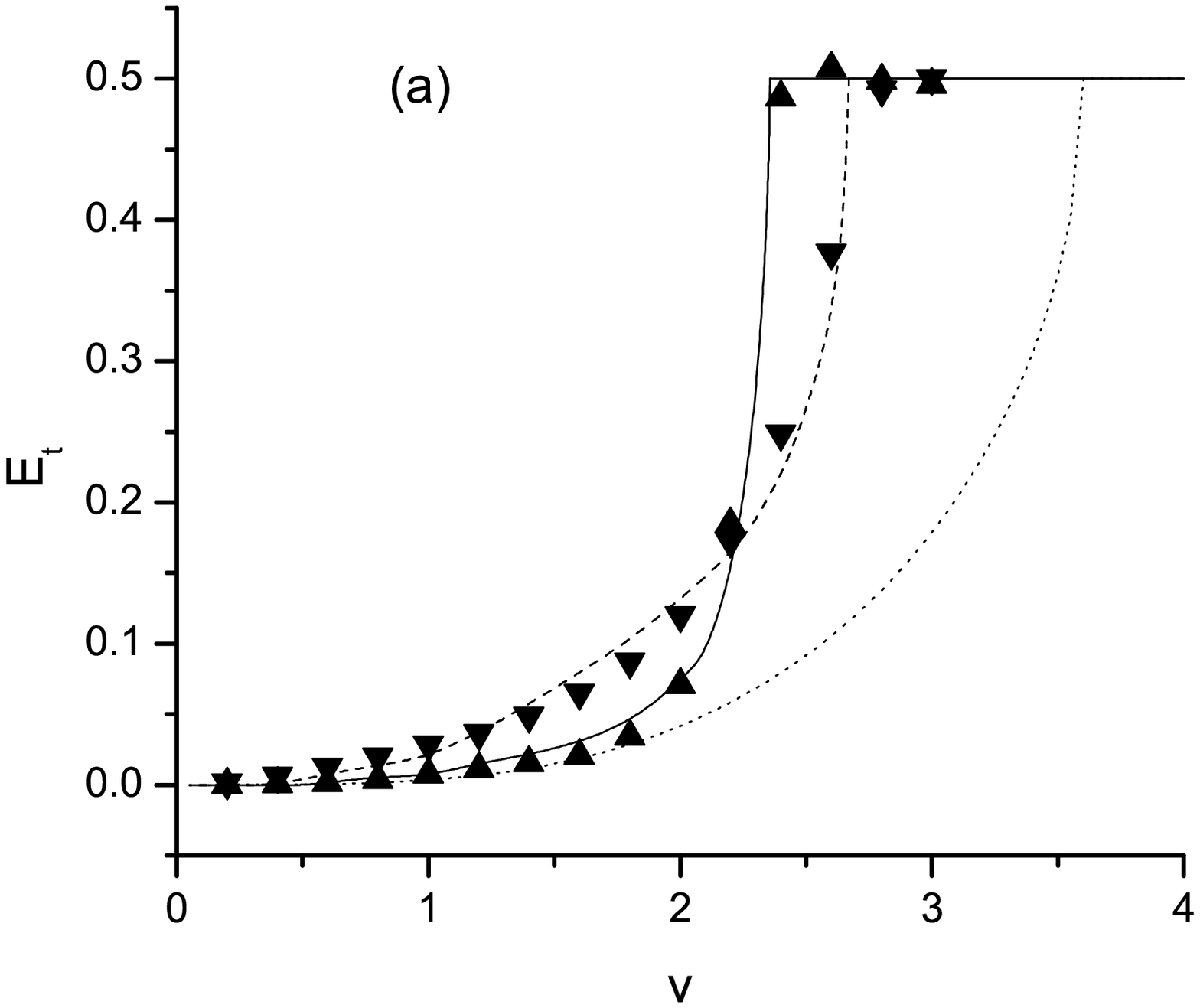}
\includegraphics[width=0.45\textwidth]{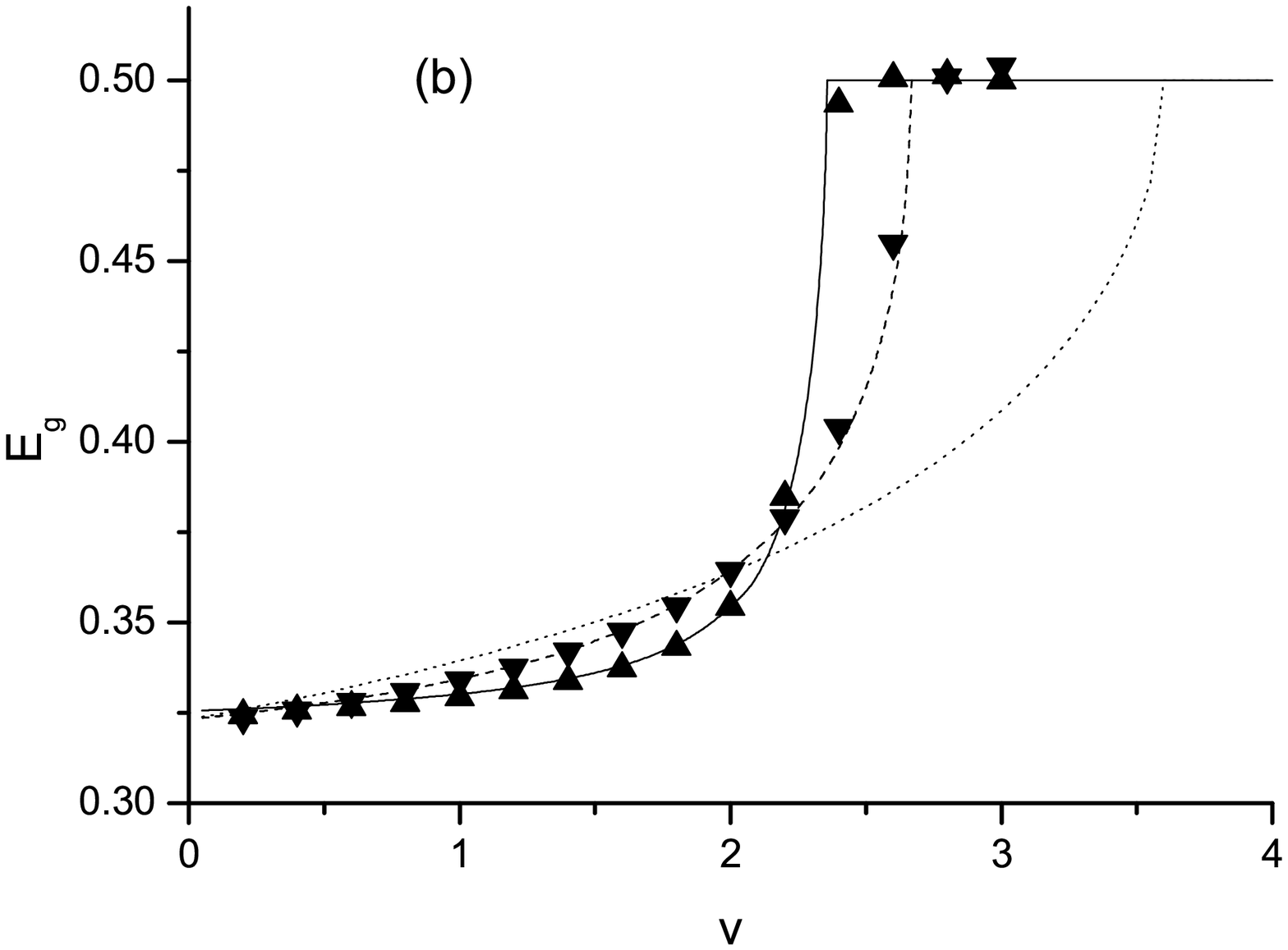}
\caption{The dependence of (a) the training error $E_t$ and
(b) generalization error $E_g$ on the learning rate $v$ for $\alpha=0.5$.
Solid lines: steady-state results of the cavity method for $\lambda=0.4$.
Dashed lines: steady-state results of the cavity method for $\lambda=0.8$.
Corresponding symbols: simulations averaged over 100 samples
with $N=1,000$ and $t=20$.
Dotted lines: the mean-force approximation for $\lambda=0.8$.}
\label{fig:res1}
\end{center}
\end{figure}

In Fig.~3, we also present the results obtained by the mean-force 
approximation adopted in Ref.~\cite{Heimel:Coolen}. As we have shown above, in 
steady state, the sequence noise due to the 
random drawing of examples at each learning step is equivalent to the 
external dynamical noise in batching learning. Thus, 
the dynamical variables will fluctuate around their temporal average even 
without other external noises.
As a result, the mean-force approximation is only
valid when the learning rate and the sequence noise is small. As shown in
Fig.~3, it has an increasing discrepancy with simulations when $v$ becomes 
large. The critical learning rate $v_c=2(1+\lambda)$ estimated by the 
mean-force approximation is larger than the simulation result.
This discrepancy can be attributed
to the omission of the force fluctuations therein.

An important question is whether on-line learning
can perform as well as batch learning \cite{Amari,Opper}. We have proposed an 
averaged strategy  in the context of batch learning, predicting that dynamical
noise can be averaged over to yield performances approaching noiseless 
learning \cite{Wong:Li:Tong}. Since sequence noise in on-line learning has 
similar effects as dynamical noise, we adopt the same strategy to improve 
performance of on-line learning.  

\begin{figure}
\begin{center}
\includegraphics[width=0.45\textwidth]{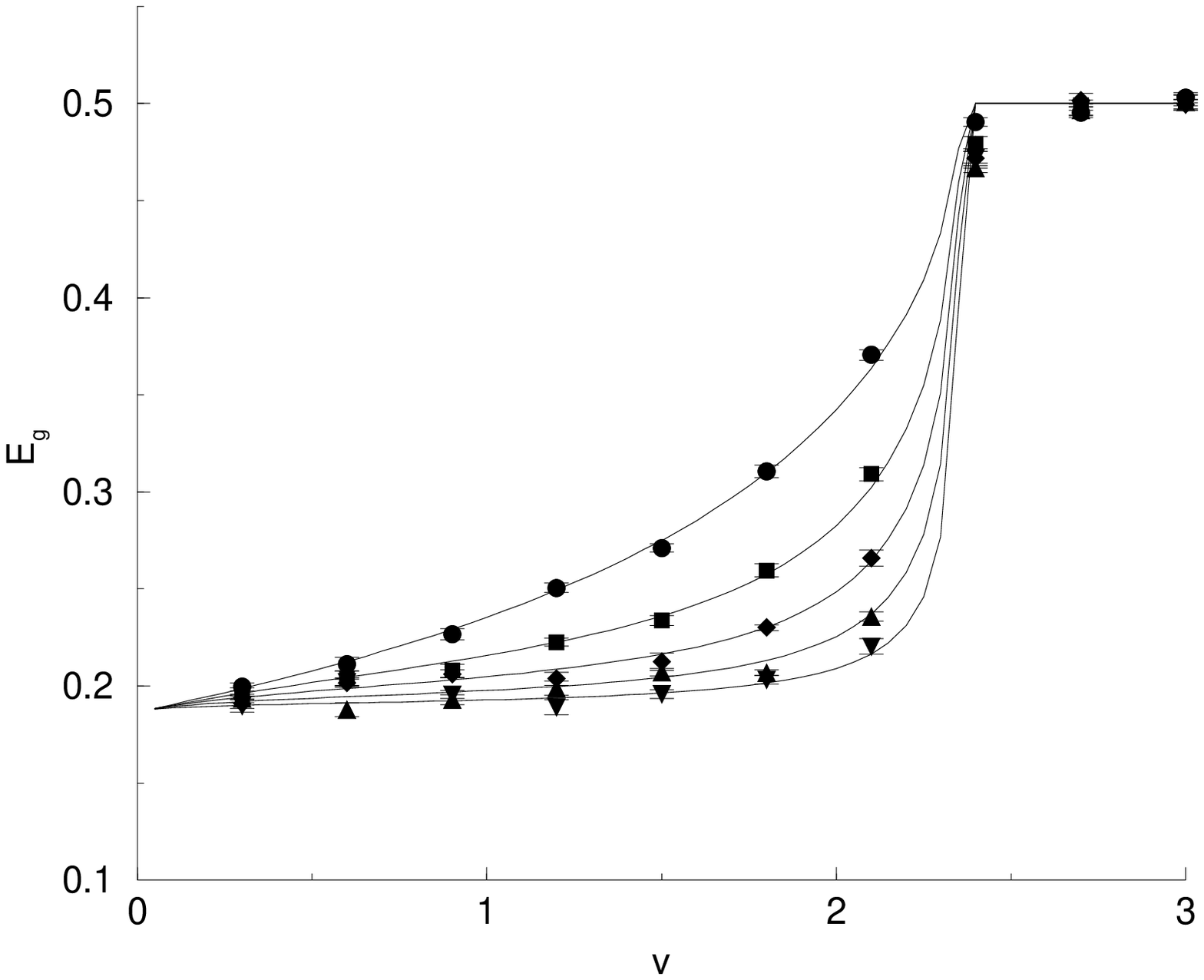}
\caption{The steady-state generalization error $E_g$
averaged over different monitoring times at $\alpha=2$ and $\lambda=0.2$.
Symbols: simulations averaged over 10 samples
with $N=1000, \tau=0$ (\ding{108}), 2 (\ding{110}),
5 (\ding{117}), 10 (\ding{115}) and 20 (\ding{116}). 
Corresponding lines: theory.}
\label{fig:res2}
\end{center}
\end{figure}

In average learning, we first wait for the system to settle to the steady 
state, and then monitor the student weight vector for an extended period of 
time. We use the weight vector averaged over the monitoring period $\tau$,  
${\bar{\boldsymbol J}}=\lim_{t\to+\infty}\frac{1}{\tau}\int_t^{t+\tau}
{\boldsymbol J}(t'){\rm d}t'$, as the estimated student vector.
The average weight amplitude is given by 
\begin{equation}
	\bar C(\tau)=\lim_{t\to\infty}\frac{1}{\tau}\int_t^{t+\tau}{\rm d}t'
			C(t,t+t').
\end{equation}
Using Eq.~(\ref{C}), we obtain
\begin{eqnarray}
	\bar C(\tau)&=&\int{\rm d}x\rho(x)\frac{x-v\lambda}{x^2}\left[
	\frac{bv}{\alpha}+a^2\left(x-v\lambda-\frac{v}{\alpha}\right)\right]
			\nonumber\\
        &&+\frac{v}{2}\langle F^2\rangle\int{\rm d}x\rho(x)
        \left(\frac{x-v\lambda}{x}\right)
        \left(\frac{e^{-x\tau}-1+x\tau}{x^2\tau^2/2}\right),
\end{eqnarray}
where $\langle F^2\rangle$ is given by Eq.~(\ref{F-2}).
We note that when $\tau$ becomes vary large,
the contribution due to force fluctuations vanishes,
making $\bar C(\tau)$ approaching the result for batch learning.
Hence, the {\it average learning strategy}
yields a generalization performance as good as that of the batch mode,
{\it independent of the learning rate,
as long as it is below the critical value for divergent learning}.
Figure 4 shows that the generalization error of on-line learning 
is equal to that of the batch mode at $v=0$,
and gradually grows larger when $v$ increases.
When the monitoring time $\tau$ increases,
there is an impressive reduction of the generalization error
compared with its instantaneous values.
The longer the monitoring time $\tau$,
the smaller generalization error.
In the limit of very long monitoring time,
the generalization ability becomes equal to that of the batch mode
for all values of $v$ below $v_c$,
and jumps discontinuously to divergence at $v_c$.

\section{Conclusion}
\label{sec:concl}

We have analysed the dynamics of on-line learning 
with restricted sets of examples, 
which are randomly recycled.
Using the cavity approach,
we can derive equations for the macroscopic parameters 
describing the learning dynamics.
They are solvable for linear rules such as the Adaline rule, 
yielding results which agree with simulations.
We also show that the student in on-line mode can learn 
as well as that in batch mode, 
after it is averaged over a monitoring period at the steady state.

Our work represents a step forward from two recent treatments
\cite{Barber:Sollich,Heimel:Coolen}.
Compared with \cite{Barber:Sollich},
we have found that a functional relationship exists
between the generic and cavity activations,
and made explicit the distribution
of activations, which is a superposition of Gaussians for Adaline rule and
each weighted by a Poisson distribution.
As a result, our framework can be used to analyse the training error $E_t$.
More importantly, it has the potential to be extended
to analyse nonlinear and multilayer networks.
Compared with \cite{Heimel:Coolen},
we have based our analysis on a more physical picture.
For example, we have made explicit the difference
between the active and passive averages
involving the activation of an example and its learning
force at a previous instant.
This enables us to analyse correctly the network behavior
at large learning rates,
where the mean-force approximation does not apply.
The physical insights will be useful
when analytical approximation schemes are devised
for more complex networks.

We have achieved our objective of benchmarking the cavity approach 
using Adaline learning. 
The next step will be to extend the method to more complicated situations 
such as nonlinear and multilayer networks. 
We may devise efficient Monte Carlo sampling procedures to solve numerically
the equations for the Green's functions, the teacher-student correlations and 
the student autocorrelations, making use of the Poisson distribution of the 
learning events for a single example, the Gaussian distribution of the cavity
activations, and their causal relations with the generic student activations.

The cavity analysis of linear networks is also the foundation of approximate 
descriptions of the stationary behavior of nonlinear and multilayer
networks. One may describe the steady state by fluctuations about
an averaged state. The fluctuations can be approximated by linear deviations
which can be analysed by the cavity approach analogous to the Adaline 
benchmark. Present work in progress is moving along this direction. 

Recently, various approximation schemes have been proposed in different 
learning regimes of complicated networks 
\cite{Coolen:Saad,Barber:Sollich,Amari}, yielding results with varying degree 
of success. For example, the conditionally Gaussian approximation cannot 
capture the sharp peaks in the activation distribution developed at the late
learning stage of nonlinear rules (e.g. Adatron learning). On the other hand,
from the perspectives of the cavity method, the nonlinear mapping from the 
Gaussian cavity activations to the generic activations can lead to sharp peaks 
in a natural way \cite{EPL}.
It is hoped that the cavity framework can provide useful
insights to improved approximations in the future.

\begin{acknowledgments}
We thank Ton Coolen for meaningful communications.
This work is supported by the Research Grant Council of Hong Kong 
(HKUST6157/99P and HKUST6153/01P).
\end{acknowledgments}

\appendix

\section{Sequence average of the student activation}
\label{sec:seq} 

According to the Eq.~(\ref{seq-avg}) and Eq.~(\ref{evolve}), the sequence 
average of the student activation of an example is
\begin{equation}                              \label{x-avg}
     \langle x(t)\rangle_\sigma=h(t)+v\sum_{m=1}^\infty
\frac{e^{-\frac{t}{\alpha}}}{\alpha^m}\int_0^t\!{\rm d}t_m\cdots
\int_0^{t_2}\!{\rm d}t_1\sum_{r=1}^m G(t,t_r)F(t_r). 
\end{equation}
Reordering the summations and the integrations before and after $t_r$,
\begin{eqnarray}
     \langle x(t)\rangle_\sigma&=&h(t)	\nonumber \\
	&&+v\sum_{m=1}^\infty	
     \frac{e^{-\frac{t}{\alpha}}}{\alpha^m}\sum_{r=1}^m\int_0^t{\rm d}t_r
	\!\int_{t_r}^t\!{\rm d}t_{r-1}\cdots\int_{t_2}^{t}{\rm d}t_1  
	\left[\!\int_0^{t_r}\!{\rm d}t_{r+1}\cdots
	\int_0^{t_{m-1}}{\rm d}t_mG(t,t_r)F(t_r)\right],  \nonumber \\
\end{eqnarray}
where we have made implicit the dependence of $F(t_r)$ on the previous 
learning instants $t_{r+1},\cdots,t_m$.
Since the integrand in the square bracket does not depend on the values 
of $t_1,\cdots,t_{r-1}$, the integration over these variables then simply 
gives the factor of $(t-t_r)^{r-1}/(r-1)!$. We further note that
\begin{equation}
   \sum_{m=1}^\infty\sum_{r=1}^m\frac{e^{-\frac{t}{\alpha}}}{\alpha^m}\cdot
	\equiv\frac{1}{\alpha}\sum_{m-r=0}^\infty
	\frac{e^{-\frac{t_r}{\alpha}}}{\alpha^{m-r}}
	\sum_{r-1=0}^\infty\frac{e^{-\frac{t-t_r}{\alpha}}}
	{\alpha^{r-1}}\cdot,  
\end{equation}
and then derive
\begin{equation}
   \langle x(t)\rangle_\sigma=h(t)+\frac{v}{\alpha}\int_0^t{\rm d}t_r
	G(t,t_r)\left[\sum_{m-r=0}^\infty\frac{e^{-\frac{t_r}{\alpha}}}
       {\alpha^{m-r}}\int_0^{t_r}{\rm d}t_{r+1}\cdots\int_0^{t_{m-1}}{\rm d}t_m
	F(t_r)\right].
\end{equation} 
From Eq.~(\ref{seq-avg}), the summation in the square bracket is just the 
sequence average of the activation at time $t_r$, yielding Eq.~(\ref{x-h}).

\section{sequence average of active and passive correlations}

\label{sec:w-c} 
From Eqs.~(\ref{evolve}) and (\ref{seq-avg}), the active sequence average 
at time $s$ of the activation of example $\mu$ that is learned at time $t$ is
\begin{eqnarray}
	\langle x_\mu(s)\rangle_\sigma|_{\sigma(t)=\mu}
        =&&h_\mu(s)+v\sum_{m=0}^\infty\frac{e^{-\frac{t}{\alpha}}}{\alpha^m}
        \int_0^t{\rm d}t_1\cdots\int_0^{t_{m-1}}{\rm d}t_m
	\sum_{n=0}^\infty\frac{e^{-\frac{s-t}{\alpha}}}{\alpha^n}
        \int_t^s{\rm d}t_{m+1}\cdots
	\nonumber\\
	&&\phantom
	{h_\mu(s)+v\sum_{m=0}^\infty\frac{e^{-\frac{t}{\alpha}}}{\alpha^m}}
	\int_t^{t_{m+n-1}}{\rm d}t_{m+n}
        \left[\sum_{r=1}^{m+n}G(s,t_r)F_\mu(t_r)+G(s,t)F_\mu(t)\right].
	\nonumber\\ 
\label{active_x}
\end{eqnarray} 
Using similar arguments as in Appendix \ref{sec:seq}, the average for the
Adaline rule can be written as a self-consistent equation for $s>t$,
\begin{eqnarray}
	\langle x_\mu(s)\rangle_\sigma|_{\sigma(t)=\mu}&=&h_\mu(s)
        +\frac{v}{\alpha}\int_0^t{\rm d}t'
        G(s,t')\langle F_\mu(t')\rangle_\sigma
        +vG(s,t)\langle F_\mu(t)\rangle_\sigma	\nonumber\\
        &&+\frac{v}{\alpha}\int_t^s{\rm d}t'G(s,t')
        [\tilde y_\mu-\langle x_\mu(t')\rangle_\sigma|_{\sigma(t)=\mu}].
\label{selfcon_x}
\end{eqnarray}		
Note that $\langle x_\mu(s)\rangle_\sigma|_{\sigma(t)=\mu}=\langle x_\mu(s)
\rangle_\sigma$ when $s\leq t$. Subtracting Eq.~(\ref{x-h}) from the above 
equation, one can derive
\begin{equation}
 	\langle x_\mu(s)\rangle_\sigma|_{\sigma(t)=\mu}
	-\langle x_\mu(t)\rangle_\sigma=
	vG(s,t)\langle F_\mu(t)\rangle_\sigma
	-\frac{v}{\alpha}\int_0^s{\rm d}t'G(s,t')
	[\langle x_\mu(t')\rangle_\sigma|_{\sigma(t')=\mu}
	-\langle x_\mu(t')\rangle_\sigma].
\end{equation}
If we multiply $F_\mu(t)$ to both sides of Eq.~(\ref{evolve}) at time $s$, 
and perform sequence averages 
analogous to Eqs.~(\ref{active_x}) and (\ref{selfcon_x}), 
we obtain equations for the active and passive averages of the 
activation-force correlation,
\begin{eqnarray}
	\langle x_\mu(s)F_\mu(t)\rangle_\sigma|_{\sigma(t)=\mu}
        &=&h_\mu(s)\langle F_\mu(t)\rangle_\sigma
        +\frac{v}{\alpha}\int_0^t{\rm d}t'G(s,t')\langle F_\mu(t')
        F_\mu(t)\rangle_\sigma|_{\sigma(t')=\mu}
	\nonumber\\
	&&+vG(s,t)\langle F^2_\mu(t)\rangle_\sigma
	+\frac{v}{\alpha}\int_t^s
	{\rm d}t'G(s,t')\langle F_\mu(t')F_\mu(t)\rangle_\sigma|_{\sigma(t)
	=\mu},			\label{actvie}   \\
	\langle x_\mu(s)F_\mu(t)\rangle_\sigma
        &=&h_\mu(s)\langle F_\mu(t)\rangle_\sigma
        +\frac{v}{\alpha}\int_0^t{\rm d}t'
        G(s,t')\langle F_\mu(t')F_\mu(t)\rangle_\sigma|_{\sigma(t')=\mu}
			\nonumber\\
	&& +\frac{v}{\alpha}\int_t^s{\rm d}t'G(s,t')\langle F_\mu(t')
	F_\mu(t)\rangle_\sigma.			\label{pass} 
\end{eqnarray}
Subtracting Eq.~(\ref{actvie}) with Eq.~(\ref{pass}) leads to 
\begin{equation}
\label{w}
	W_\mu(s,t)=vG(s,t)\langle F_\mu^2(t)\rangle_\sigma-\frac{v}{\alpha}
	\int_t^s{\rm d}t'G(s,t')W_\mu(t',t).
\end{equation}
Multiplying the example Green's function $D(r,s)$ to both sides of 
Eq.~(\ref{w}), integrating over $s$, and applying Eq.~(\ref{D}),
one obtains Eq.~(\ref{act-pass}).

To obtain Eq.~(\ref{passive}), one replaces 
$\langle F_\mu(t')F_\mu(t)\rangle_\sigma|_{\sigma(t')=\mu}$
in the right hand side of Eq.~(\ref{pass}) with 
$\langle F_\mu(t')F_\mu(t)\rangle_\sigma-v\int_{t'}^t{\rm d}t_1D(t,t_1)
G(t_1,t')\langle F_\mu^2(t')\rangle_\sigma$, and arrives at
\begin{eqnarray}
	\langle x_\mu(s)F_\mu(t)\rangle_\sigma
        &=&h_\mu(s)\langle F_\mu(t)\rangle_\sigma
	+\frac{v}{\alpha}\int_0^s{\rm d}t'
        G(s,t')\langle\tilde y_\mu F_\mu(t)\rangle_\sigma
	-\frac{v}{\alpha}\int_0^s{\rm d}t'
	G(s,t')\langle x_\mu(t')F_\mu(t)\rangle_\sigma 	\nonumber \\
	&&-\frac{v^2}{\alpha}\int_{0}^t{\rm d}t_1\int_{t_1}^t{\rm d}t_2
	G(s,t_1)D(t,t_2)G(t_2,t_1)\langle F^2_\mu(t_1)\rangle_\sigma.
\end{eqnarray}
Multiplying both sides with $D_\mu(r,s)$, integrating over $s$ and applying 
Eq.~(\ref{D}), one finally reaches Eq.~(\ref{passive}).

\newpage
\bibliography{tran-st.bbl}

\end{document}